\documentclass[superscriptaddress,endfloats,amsmath,  preprint]{revtex4-1}  % for review and submission

\usepackage{graphicx}% Include figure files
\usepackage{float}
\usepackage{dcolumn}% Align table columns on decimal point
\usepackage{bm}
\usepackage{multirow,rotating} % for sideways labels in table...
\usepackage{color}
\usepackage{verbatim}
\usepackage{natbib}
\usepackage{physics}
\setcitestyle{super}

\begin{document}

\title{
Modelling Universal Order Book Dynamics in Bitcoin Market
%Order book dynamics model based on trader behavior in Bitcoin exchange market
%Stochastic growth of the order-book dynamics for BitCoin exchange market
}

\author{Fabin Shi}
\affiliation{\footnotesize{CAS Key Laboratory of Network Data Science and Technology, Institute of Computing Technology, Chinese Academy of Sciences, Beijing, China}}
\affiliation{\footnotesize{University of Chinese Academy of Sciences, Beijing 100049, China}}
\author{Nathan Aden}
\affiliation{\footnotesize{Department of Physics, University of Miami, Coral Gables, Florida 33142, USA}}
\author{Shengda Huang}
\affiliation{\footnotesize{Department of Physics, University of Miami, Coral Gables, Florida 33142, USA}}
\author{Neil Johnson}
\affiliation{\footnotesize{Physics Department, George Washington University, Washington D.C. 20052}}
\author{Xiaoqian Sun}
\affiliation{\footnotesize{CAS Key Laboratory of Network Data Science and Technology, Institute of Computing Technology, Chinese Academy of Sciences, Beijing, China}}
\author{Jinhua Gao}
\affiliation{\footnotesize{CAS Key Laboratory of Network Data Science and Technology, Institute of Computing Technology, Chinese Academy of Sciences, Beijing, China}}
\author{Li Xu}
\affiliation{\footnotesize{CAS Key Laboratory of Network Data Science and Technology, Institute of Computing Technology, Chinese Academy of Sciences, Beijing, China}}
\author{Huawei Shen}
\affiliation{\footnotesize{CAS Key Laboratory of Network Data Science and Technology, Institute of Computing Technology, Chinese Academy of Sciences, Beijing, China}}
\affiliation{\footnotesize{University of Chinese Academy of Sciences, Beijing 100049, China}}
\author{Xueqi Cheng}
\affiliation{\footnotesize{CAS Key Laboratory of Network Data Science and Technology, Institute of Computing Technology, Chinese Academy of Sciences, Beijing, China}}
\affiliation{\footnotesize{University of Chinese Academy of Sciences, Beijing 100049, China}}
\author{Chaoming Song}
\thanks{e-mail: c.song@miami.edu}
\affiliation{\footnotesize{Department of Physics, University of Miami, Coral Gables, Florida 33142, USA}}

%\institute{ \Letter  Alexw\at}
\renewcommand{\figurename}{Fig.}

\def\pmid{p_{mid}}
\def\SIn{\sigma_\pm^{in}}
\def\SOut{\sigma_\pm^{out}}
\def\SD{D_{\pm}}
\def\SH{h_{\pm}(x,t)}
%\def\@cite#1#2{$^{\mbox{\scriptsize #1\if@tempswa , #2\fi}}$}
%\newcommand{\upcite}[1]{\textsuperscript{\cite{#1}}}
%\newcommand{\upcite}[1]{\textsuperscript{\textsuperscript{\cite{#1}}}}

%\date{\today}

\begin{abstract}
Understanding the emergence of universal features such as the stylized facts in markets is a long-standing challenge that has drawn much attention from economists and physicists. Most existing models, such as stochastic volatility models, focus mainly on price changes, neglecting the complex trading dynamics.  Recently, there are increasing studies on order books,  thanks to the availability of large-scale trading datasets, aiming to understand the underlying mechanisms governing the market dynamics.  In this paper, we collect order-book datasets of Bitcoin platforms across three countries over millions of users and billions of daily turnovers.  We find a  1+1D field theory, govern by a set of KPZ-like stochastic equations, predicts precisely the order book dynamics observed in empirical data. Despite the microscopic difference of markets, we argue the proposed effective field theory captures the correct universality class of market dynamics. We also show that the model agrees with the existing stochastic volatility models at the long-wavelength limit.

\end{abstract}

\maketitle

\newpage

Understanding universal emergent properties in different markets is a long-standing challenge for both economists and physicists. As early as the 1960s, Mandelbrot~\cite{Mandelbrot1963The} pointed out that the distribution of logarithmic price return was heavy-tailed in the cotton market which, soon after, was found to hold true in numerous other markets~\cite{gopikrishnan1998inverse,plerou2008stock,gu2008empirical,gopikrishnan1999scaling}.  
Since then many many stylized facts have been observed as common across a wide range of instruments, markets, and time periods ~\cite{cont2001empirical,cont2005long,stanley2008statistical,campbell1997econometrics,chakraborti2011econophysics,gould2013limit} . This raises a fundamental question:  what are the general mechanisms in a financial market leading to these phenomena.

Existing approach esof modeling price evolution as a stochastic process to capture the volatility of a market, such as stochastic volatility (SV) models~\cite{bollerslev1986generalized, heston1993closed,beckers1980constant,hagan2002managing,ahn1999parametric}, have been met with success when attempting to tease out numerous stylized facts such as the volatility clustering and heavy-tailed price return distributions.  However since these models do not include aspects of the actual trading process, connections between these facts and human behavior remain outside their scope.  A natural extension is then to include the actions of traders as the mechanism behind creating the price by incorporating all limit orders into bid/ask order books at a given time $t$ and price $x$, and the matching price at the position $x=0$ at which buyers and sellers agree to trade. Thanks to technological advances during the past decade there are an increasing number of datasets available about order-book dynamics which provide the microscopic details of trading dynamics ~\cite{martin2013obr}. These details have been used to construct several models that attempt to bridge the gap between human behavior and market dynamics~\cite{maslov2000simple,cont2010stochastic,bouchaud2002statistical,mike2008empirical,daniels2003quantitative,smith2003statistical,kanazawa2018derivation,yura2014financial}. Bak et.al. considered orders as particles and models the movement of each particle along the price lattice using a random walk ~\cite{bak1997price}. Further work also took into account fixed limit orders and market orders that trigger transactions ~\cite{maslov2000simple}. More recently, orders were modelled as Brownian colloidal particles bumbling along in a price-fluid ~\cite{yura2014financial} .  However the common approach in these models is the discretization of price which has the potential to obfuscate the behavior/market connection with details about how orders are transacted in the specific market analyzed.  In our model we ignore some of these details by smoothing out the limit order price axis into a continuous spacial dimension.  Along with a continuous time axis, we propose a 1+1D field theory to explain some of the stylized facts as resulting directly from the tendencies of traders in a limit order based market.

\section{Analyzing and modeling the order-book dynamics}
Despite there being several studies based on the order-book datasets for varies securities, these datasets are often limited by quantity, time span, and accessibility. The novelty of Bitcoin however lies in the decentralized nature of how transactions are executed. Trades involving BTC are only recognized as valid once they have been communally mined into the publicly available ledger, which is known as Blockchain.  This intrinsic market data availability has lead to extensive study since its inception in 2009.  The first Bitcoin exchanges emerged in 2010 providing a uniquely public look into the mechanics of exchange trading including orderbook dynamics.  Some early analyses of this data focused primarily on standard financial methods to compare Bitcoin to normal currencies \cite{raventos2012bitcoin}.  Later works explored price prediction and stability analysis \cite{shah2014bayesian, donier2015markets, donier2015million}. More recently, Bitcoin has entered the public discourse by exploding in value throughout 2017 and then bursting soon after in early 2018 thereafter continuing to rise and fall in diminishing motions, seemingly approaching a stable value.  This long and varied public economic history makes it an ideal candidate on which to test our model.

We use three datasets collected through different online Bitcoin trading platforms: {\bf (i)} OKCoin was the largest Bitcoin exchange platform in China, consisting of millions of users and billions of turnovers per day until being shut down in 2017 due to government policy. We collected order-book data from OKCoin from Nov. 3rd, 2016 to Jul. 28th, 2017 (with an unfortunate gap from Jan. 4th, 2017 to Mar. 1st, 2017 due to machine failure).  Since OKCoin introduced an additional transaction fee on each order after Jan. 24, 2017, we decided to split the data in two: Nov. 3rd, 2016 to Jan. 4th, 2017  (OKCoin1) and Mar. 1st, 2017 to Jul. 28th, 2017 (OKCoin2). {\bf (ii)} We also collected data from BTC-e, one of the largest Bitcoin trading platforms headquartered in Russia, from May 3rd, 2017 to Jul. 26th, 2017.  {\bf (iii)}  And lastly from Coinbase, a US-based Bitcoin trading platform, from Jan. 23rd, 2018 to Apr. 18th, 2018. The order-book datasets collected for each of these three domains record the profiles of the bid (limit buy) and ask (limit sell) orders every few seconds during the stated observation period. We are unable to track the instantaneous change of each order.  Nevertheless, for OKCoin1 we also collected the market order transaction data per second by recording the total number of market orders which are higher/lower than the best price (bid/ask) and immediately match to one or more active orders upon arrival.

    \begin{figure}[H]
        \centering
        \includegraphics[width=\textwidth,height=17cm]{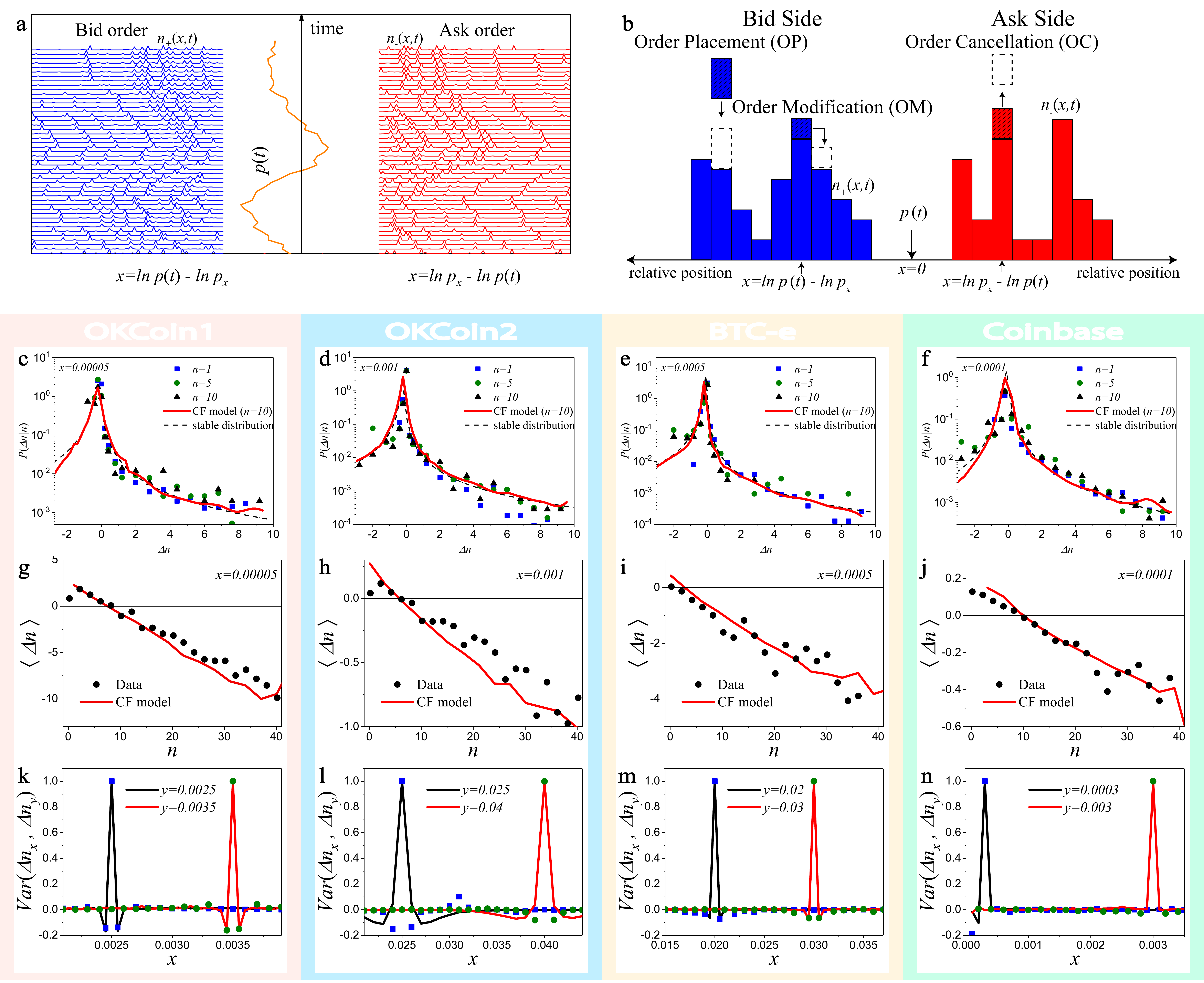}\\
        \caption{{\bf Analysis and modelling the three types of order-book operations in different Bitcoin markets.} {\bfseries{a}}) A typical ask/bid order-book profile.  {\bfseries{b}}) The schematic description of three order-book operations. {\bfseries{c--f}}) The  conditional distribution,  $P(\Delta n|n)$ for {\bfseries c}) OKCoin1, {\bfseries d}) OKCoin2, {\bfseries e} BTC-e and {\bfseries f}) Coinbase. Dots denote measurements from data and lines are measurements from simulation. {\bfseries{g--j}}) The change of order volume  $\Delta n$ versus order volume $n$ for {\bfseries g}) OKCoin1, {\bfseries h}) OKCoin2, {\bfseries i} BTC-e and {\bfseries j}) Coinbase. Dots denote measurements from data and lines are measurements from simulation. {\bfseries{k--n}}) The correlation $\langle \Delta n_x,\Delta n_y\rangle$~(the correlation between the change of order volume at different positions) versus position $x$ for {\bfseries k}) OKCoin1, {\bfseries l}) OKCoin2, {\bfseries m}) BTC-e and {\bfseries n}) Coinbase. Dots denote measurements from data and lines are measurements from simulation.} \label{fig1}
    \end{figure}

We introduce a $1+1 D$ continuous field (CF) model to explain the dynamics found in the bid/ask order volumes, $n_+\qty(x,t)$ and $n_-\qty(x,t)$.  The spatial dimension $x\equiv \pm \ln{p\qty(t)}\mp \ln{p_x}\geq 0$ is the logarithmic distance between an order price $p_x$ and the trading price $p\qty(t)$ with the two signs correspond to the bid/ask axes respectively for the notational convenience of keeping $x$ positive. Fig.~\ref{fig1}{{a}} demonstrates a typical bid/ask order-book profile over time.  
Figure~\ref{fig1}{{c--f}}  plots the distribution of the order volume change among bids, $\Delta n_+\qty(x,t)\equiv n_+\qty(x,t)-n_+\qty(x,t-\Delta t)$, for a fixed $x$ and $n$ and various values of $\Delta t$, revealing a fat-tailed nature for both positive and negative tails. Similar results observed among the ask side for $\Delta n_-$. Any change in the volume of orders away from the $x=0$ boundary must come from one of three possible order-book operations, i) order placement (OP), ii) order cancellation (OC), and iii) order modification (OM), as  illustrated in Fig.~\ref{fig1}{{b}}. We model these three operations as follows:

{\textbf {(1)}}{\it Order Placement}: Traders place a new order on top of previous orders at some price $x\neq 0$.   It  suggests that in the continuous case we can model the change in order volume due to order placement, notated as $dn_\pm^{OP}(x,t)$ as
    \begin{equation}\label{eq:OP}
         dn_\pm^{OP}(x,t)=\SIn(x)\xi_\pm(x,t)dt,
    \end{equation}
where $\xi_\pm(x,t)$ is continuous set of random variables satisfying some one-sided stable distribution. We find that we must allow the scale parameter $\SIn$ to depend on the position $x$. This general ingredient of order-book dynamics has been found in both the Paris Stock Exchange~\cite{bouchaud2002statistical} and the London Stock Exchange~\cite{zovko2002power}.

{\textbf {(2)}} {\it Order Cancellation}: Traders cancel orders which they have placed previously. In Fig.~\ref{fig1}{{g--j}}, we have plotted the time averaged change of order volume at some fixed $x$, $\ev{\Delta n_+\qty(x,t)}_t$ against the current order volume. Unlike the Order Placement (\ref{eq:OP}) where changes are independent of $n$,  we see a linear dependence consistent with an existing study~\cite{challet2001analyzing} from which we can intuit the form of the order cancellation term to be
    \begin{equation}\label{eq:OC}
        dn_\pm^{OC}(x,t)=-\SOut(x)n_\pm(x,t)\zeta_\pm(x,t)dt.
    \end{equation}
The scale parameter $\SOut$, similar to $\SIn$, depends on the current position $x$ and again $\zeta_\pm(x,t)$ is a random variable satisfying the same stable distribution above.

{\textbf{(3)}} {\it Order Modification}: Traders change the price of orders that they own. Empirically there exists a negative correlation between $\Delta n_+\qty(x,t)$ at different positions in Fig.~\ref{fig1}{{k--n}} suggesting that the order modification operation can be modeled as a diffusion process along the order-books. Therefore, the order modification term is
    \begin{equation}\label {eq:OM}
         dn_\pm^{OM}(x,t)=\frac{\partial^2}{\partial x^2}\SD(x)n_{\pm}(x,t)dt,
    \end{equation}
where the diffusion rate $\SD(x)$ depends on the position in general. It is possible that the negative correlation we observed is due to a combination of order modification and the correlated behaviors of adding/removing orders, perhaps through different users. As an effective field model such microscopic differences are effectively the same and all captured by the diffusion term (see Supplementary Section S2 for a direct validation of  ~\eqref{eq:OM} using an additional dataset).

Directly from the chain rule we obtain
    \begin{equation} \label{eq:ChainRule}
        \frac{d n_\pm(x, t)}{dt} = \frac{ \partial n_\pm(x,t) }{\partial t} \pm \frac{ \partial n_\pm(x,t) }{\partial x} v(t),
    \end{equation}
where $v(t) \equiv d {\ln p(t)}/dt $ is the velocity of logarithmic price.  The total derivative would then be simply the sum of the effects of order operations determined above~\eqref{eq:OP}--\eqref{eq:OM} leading to our first stochastic differential equation
    \begin{equation}\label{eq:model}
        \frac{\partial n_{\pm}(x,t)}{\partial t}=\frac{\partial^2\SD(x)n_{\pm}(x,t)}{\partial x^2}\mp v(t)\frac{\partial n_{\pm}(x,t)}{\partial x}+\SIn(x)\xi_\pm(x,t)-\SOut(x)n_\pm(x,t)\zeta_\pm(x,t).
    \end{equation}
Unlike limit orders, when market orders are placed they are set to execute immediately at the trading price -- even before limit orders momentarily existing at the $x=0$ boundary.  Therefore the discrepancy in these orders placed in a short period of time, denoted $J\qty(v) \equiv \Delta n^{MO}_+\qty(v,t)-\Delta n^{MO}_-\qty(v,t)$  controls the flow of orders through the $x=0$ boundary meaning a positive excess would indicate more buyers than sellers so the discrepancy would begin depleting the reservoir of ask limit orders and vice-versa.  Applying the continuity equation gives the rate of change of the total volume  in $n_{\pm}\qty(x,t)$ as $\pdv{x}\qty(D_\pm\qty(0,t)n_{\pm}\qty(0,t))\mp v\qty(t)n_{\pm}\qty(0,t)$ which must be conserved by the market orders leading to
    \begin{equation}\label{eq:velocity}
        v(t)=\frac{1}{n_0(t)}\left[J(v,t) +\frac{\partial D_-n_-}{\partial x}(0,t)-\frac{\partial D_+n_+}{\partial x}(0,t)\right],
    \end{equation}
where $n_0(t)=n_+(0,t)+n_-(0,t)$.  Equations.~\eqref{eq:model}--\eqref{eq:velocity} give a complete description of our CF model which exhibits the relationship between order placement, order cancellation, order modification, and price change.

\begin{figure}[H]
        \centering
        \includegraphics[width=\textwidth]{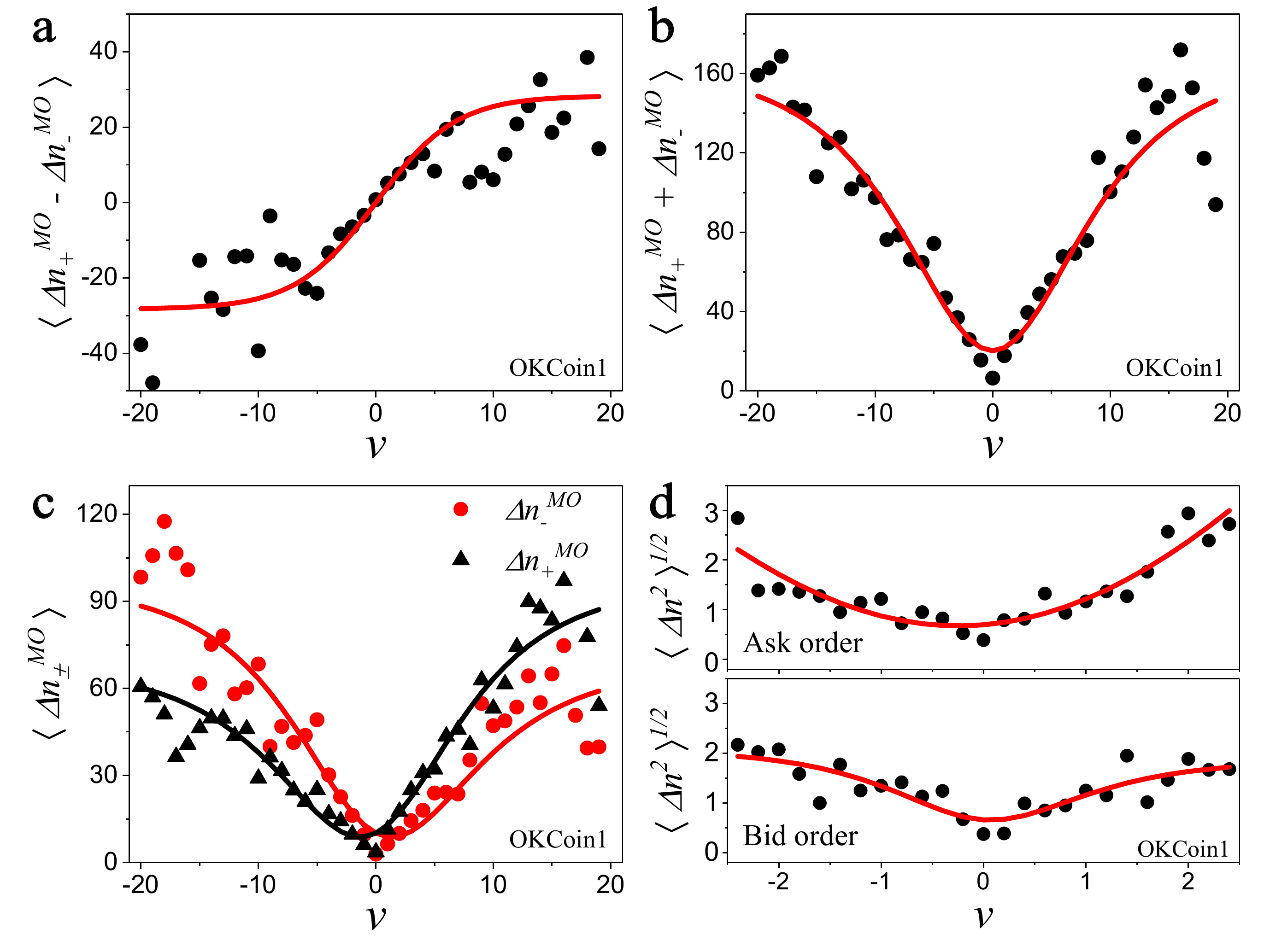}\\
        \caption{{\bf The traders' reaction to velocity.} {\bfseries{a}}) The discrepancy of market order $\Delta n_+^{MO}-\Delta n_-^{MO}$ versus velocity $v$ in OKCoin1. Dots  denote measurements from data, whereas the curve is a guide to the eye, following $\Delta n_+^{MO}-\Delta n_-^{MO}\propto k_0\tanh(v/v_0)$. {\bfseries{b}}) The market order volume $\Delta n_+^{MO}+\Delta n_-^{MO}$ versus velocity $v$ in OKCoin1. Dots  denote measurements from data, whereas the curve is a guide to the eye, following $\Delta n_+^{MO}+\Delta n_-^{MO}\propto k_{\infty}-k_{\infty}\sech(v/v_0)$. {\bfseries{c}}) The market order $\Delta n_\pm^{MO}$ versus velocity $v$ in OKCoin1. Dots  denote measurements from data, whereas the curve is a guide to the eye, following $\Delta n_\pm^{MO}\propto [\pm k_0\tanh(v/v_0)+k_\infty-k_{1}\sech{(v/v_0)}]v_0$. {\bfseries{d}}) The root mean square of $\Delta n$ versus normalized $v$ in OKCoin1. Dots  denote measurements from data, whereas the line is a guide to the eye, following $\langle \Delta n\rangle^{1/2}\propto [\pm k_0^{'}(x)\tanh(v/v_0^{'}(x))+k_\infty^{'}(x)-k_{1}^{'}(x)\sech{(v/v_0^{'}(x))}]v_0^{'}(x)$.}\label{fig2}
    \end{figure}

From here we describe two important aspects of the traders' reactions to velocity of the price, $J\qty(v,t)$. The first is the influence of trend-following.  The intuition being that the traders will try to follow the changing price \emph{e.g.} that traders would prefer placing bid orders as the price is increasing and ask orders as it is decreasing. In Fig.~\ref{fig2}{{a}}, we observe exactly this: $J\qty(v,t)$ is the linear response to $v$ for small velocity but also saturates at high speeds. The work done by Kanazawa~\cite{kanazawa2018derivation} suggests that this curve approximately follows a hyperbolic tangent. Thus we set $J\propto\tanh({v}/{v_0})$, fitting the empirical data well. The other is the influence of market activity. When the market is moving at high speeds in either direction, it seems to cause more activity among the traders. In Fig.~\ref{fig2}{{b}}, the total change in market order volume over a small time-step $\Delta n_+^{MO}+\Delta n_-^{MO}$ is observed to increase as the magnitude of the velocity grows, verifying the existence of this influence. We chose a natural fit to this data using $\Delta n_+^{MO}+\Delta n_-^{MO}\propto 1- \sech({v}/{v_0})$. These two equations combine to describe the behavior of market orders~(Fig.~\ref{fig2}{{c}}),
    \begin{equation}\label{eq:MO}
        \Delta n^{MO}_{\pm}=[\pm k_0\tanh({v}/{v_0})+k_\infty-k_{1}\sech{({v}/{v_0})}]v_0.
    \end{equation}    
%where we see that $\Delta n^{MO}_{\pm}\qty(v\gg v_0) \sim v\qty[k_\infty \pm k_0]$ and $\Delta n^{MO}_{\pm}\qty(v\ll v_0) \sim v \qty[k_\infty - k_1]$ gives an intuitive understanding of what the $k$ parameters are fitting.

We also analyzed the rms change in the total number of limit orders over short period of time and found that it too approximately follows equation~\eqref{eq:MO} according to~(Fig.~\ref{fig2}{{d}}).  It is then reasonable to believe that the traders' reactions to the movement of the trading price at any $x$ should mirror in form that of the reaction seen in market order activity.  We propose that the limit order placement activity function is of the form $\sigma^{in}(x,v)=[ k_0^{in}(x)\tanh({v}/{v_0^{in}(x)})+k_\infty^{in}(x)-k_{1}^{in}(x)\sech{({v}/{v_0^{in}(x)})}]v_0^{in}(x)$ where to avoid cluttering the notation we have left off the $\pm$ subscripts.

\section{Model Predictions}
To test the validity of our model, we conduct some simulations of the order-book dynamics~(Supplementary Section S1) and compare the simulation results with empirical data in the OKCoin1, OKCoin2, BTC-e, and Coinbase datasets. We first indirectly provide evidence supporting the validity of the form of the three trader operations that we have included in the model. A consideration of the diffusion-less ($D_{\pm}\qty(x)=0$) and point process $J = 0$ limits of our model consisting only of traders placing and canceling orders at random leads to a linear relationship between $\ev{\Delta n_{\pm}\qty(x,t)}_t$ and $\ev{n_{\pm}\qty(x,t)}_t$ which is verified in ~(Fig.~\ref{fig1}{{g--j}}) since empirically the contributions of the diffusion term were small on time scales where the velocity doesn't change very much.  We also see justification for the heavy tails of $\xi_{\pm}\qty(x,t)$ and $\zeta_{\pm}\qty(x,t)$ in the heavy tail observed on the distribution for $\Delta n_{+}$~(Fig.~\ref{fig1}{{c--f}}).  The final row of figures~(Fig.~\ref{fig1}{{k--n}}) show the classic signs of the negative rebounds on either side of the self-correlating spike common to diffusion processes with a more detailed analysis given in the supplementary materials~(Supplementary Section S2).
    \begin{figure}[H]
        \centering
        \includegraphics[width=\textwidth]{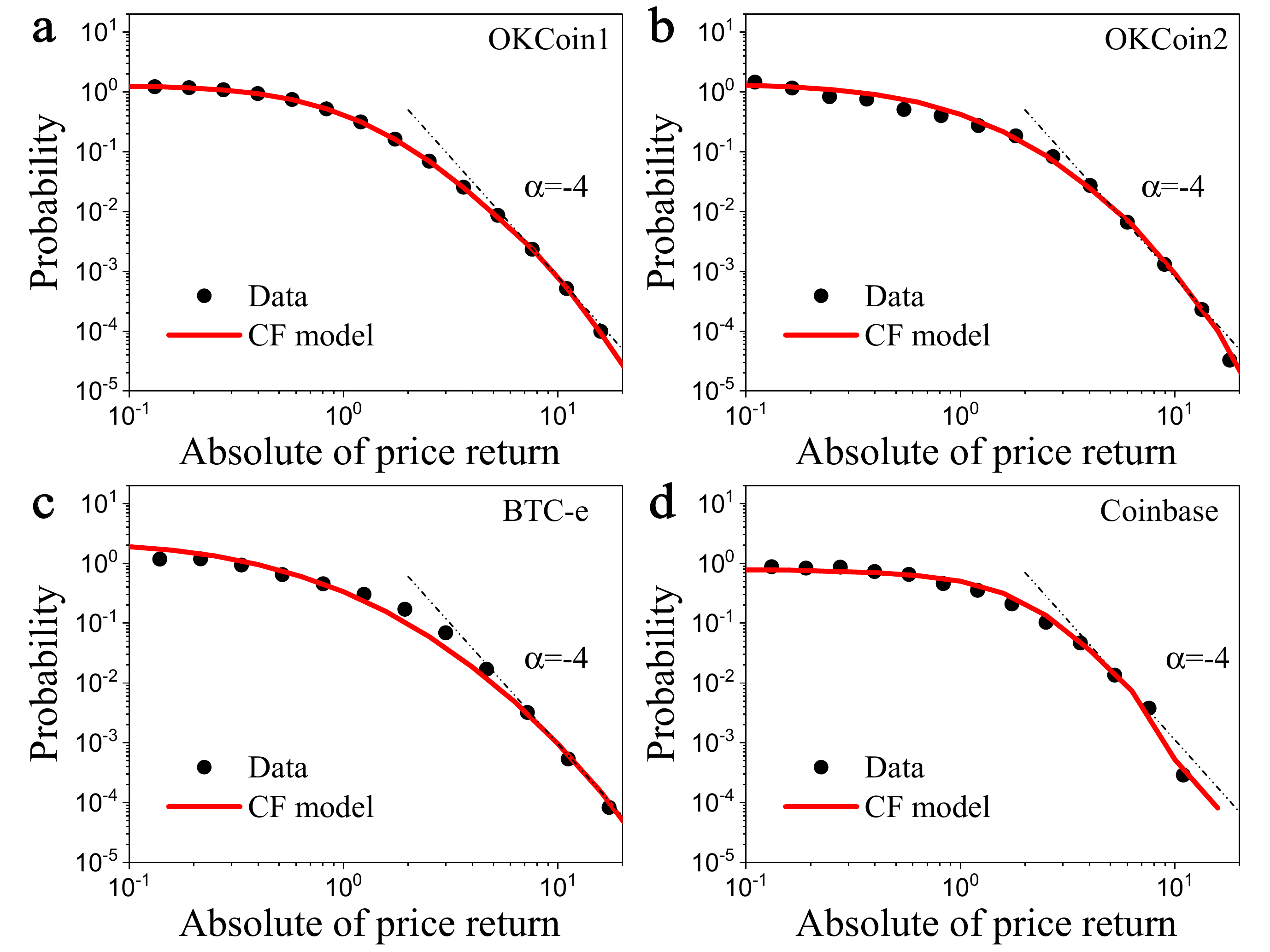}\\
        \caption{{\bf {The distribution of absolute value of the normalized price return in different Bitcoin market.}} The probability distribution of absolute value of the normalized price return  for {\bfseries a}) OKCoin1, {\bfseries b}) OKCoin2, {\bfseries c}) BTC-e and {\bfseries d}) Coinbase. Dots denote measurements from data, lines are measurements from simulation.  The dot dash, shown as a guide to the eye, represents a power-law decay with exponent $\alpha=-4$.
         }\label{fig3}
    \end{figure}

Moreover, we measure the distribution of the absolute value of instantaneous price return, which is important for understanding the market, quantifying risk, and optimizing portfolios~\cite{bouchaud2003theory,johnson2003financial}. Because it is defined as the logarithmic ratio of the price before and after the smallest discernible unit of time in the market (tic, $\tau$) the price return is equivalent to the velocity times the tic $\abs{v\qty(t)\tau}$. In Fig.~\ref{fig3}{{a--d}}, we plot the pdf for the price return normalized to absorb the effect of tic size which is of course irrelevant in our continuous model.  We show that the heavy tail of this distribution decays with an exponent of $\alpha\approx -4$ in agreement with our theory which reproduces the well known cubic (quartic) law of returns found in the ccdf (pdf) for many different financial markets~\cite{gopikrishnan1998inverse,gopikrishnan1999scaling,chakraborti2011econophysics,gould2013limit}.  As it will be shown, the method in which our model predicts this exponent is very general suggesting that this mechanism is a sufficient explanation for this universality class independent of Bitcoin specific market details.

Simulations reveal the diffusion terms in equation~\ref{eq:velocity} to be negligible in the influence on price movement allowing $v\qty(t+\tau)\approx{J\qty(v\qty(t))}/{n_0}\qty(t2S)$ where care has been taken in the writing the correct time dependence.  Thus we construct the infinitesimal for velocity as $dv \approx \frac{J\qty(v)}{n_0}-v$ where every quantity is now evaluated at the same time.  We can construct the Fokker-Planck equation for the distribution of the returns by writing the it$\hat{\text{o}}$ SDE $dv = \mu\qty( v)dt+\sigma\qty(v) dW_t$.  We then measure the drift and diffusion coefficients by finding the relationships $\ev{\Delta v}= -v$ so that $\mu\qty(v) = \dv{t}\ev{v} \approx \frac{1}{\tau}\ev{\Delta v} = -\frac{v}{\tau}$ and $\ev{\text{Var}\qty(v)}\approx 0$ when $v=0$ so that $k_1\approx k_\infty$ and $\sigma^2\qty(v)=\frac{v^2_0}{n^2_0\tau^2}\qty[k_0^2\tanh[2](\frac{v}{v_0})+\qty(k_\infty-k_1 \sech[2](\frac{v}{v_0}))]$~(Supplementary Section S3).  We then use the Fokker-Planck equation,
\begin{equation}\label{eq:differential_velocity}
\frac{\partial}{\partial t}p(v)=-\frac{\partial}{\partial v}[\mu(v)p(v)]+\frac{\partial^2}{\partial v^2}[\frac{\sigma^2(v)}{2}p(v)].
\end{equation}
to solve for the stable solution
\begin{equation}\label{eq:probability_velocity}
p(v)\propto\frac{2}{\sigma^2(v)}e^{2\int{\frac{\mu(v)}{\sigma^2(v)}dv}}.
\end{equation}
which is the general form of the price return distribution.  A summary of solutions to this equation are given below
\begin{align*}
	p\qty(v) &\propto \text{exp}\qty(-\frac{n_0^2v^2}{v_0^2\qty(k_\infty-k_1)^2})  &       &\abs{v}\to 0\\
	p\qty(v) &\propto v^{-2-2k_0^{-2}n_0^2}                & 0\ll  &\abs{v}\ll v_0 \\
	p\qty(v) &\propto \text{exp}\qty(-\frac{n_0^2v^2}{v_0^2\qty(k_\infty+k_1)^2})  & v_0\ll&\abs{v}
\end{align*}
In the regime where the power law dominates we find that $n_0\approx k_0$ in the OKCoin1 dataset~(Supplementary Section S3) which gives a power of $-4$.
    \begin{figure}[H]
        \centering
        \includegraphics[width=\textwidth]{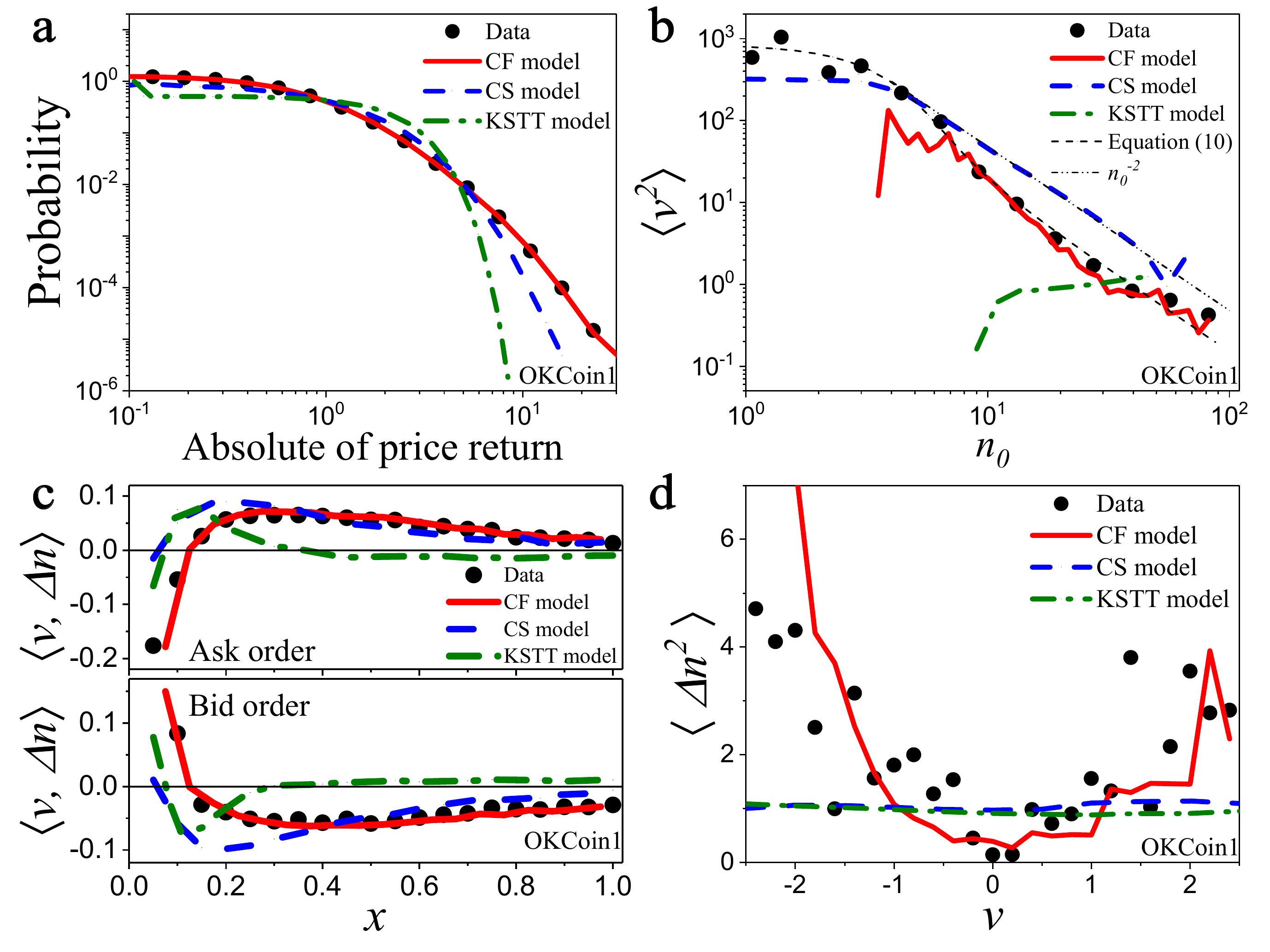}\\
        \caption{{\bf The properties of order-book dynamic in OKCoin1.} {\bfseries{a}}) The probability distribution of the absolute value of normalized price return in OKCoin1. Dots denote measurements from data and lines are measurements from different models. {\bfseries{b}}) The variance of velocity $\langle v^2 \rangle$ versus order volume $n_0$ in OKCoin1. Dots denote measurements from data and lines are measurements from different models. {\bfseries{c}}) The correlation $\langle v,\Delta n\rangle$ (correlation between the change of order volume $\Delta n$ and velocity $v$) versus position $x$ in OKCoin1. Dots denote measurements from data and lines are measurements from different models. {\bfseries{d}}) The root mean square of the change of order volume $\langle \Delta n^2\rangle^{1/2}$  versus velocity $v$ in OKCoin1. Dots denote measurements from data and lines are measurements from different models.
        }\label{fig4}
    \end{figure}

We also compare our model to two existing models -- CS~\cite{challet2001analyzing} and KSTT~\cite{kanazawa2018derivation}.  The CS model deals with specific trader behavior in allowing for the placement and cancellation of orders whereas the KSTT model focuses on the traders' reaction to changing price.  However neither produce a price return distribution with the appropriate universal exponent~(Fig.~\ref{fig4}{a}) since in both models the variance of the change in velocity is independent of velocity which, according to equation~\eqref{eq:probability_velocity}, implies that the distribution of price return follows a Gaussian.

In addition to price return, we also verify some other useful quantities from the model with our data. The relation between second moment of the velocity $\langle v^2\rangle$ and market order volume $n_0$ is calculated with an expectation with respect to the conditional distribution on $n_0$ so we have
    \begin{equation}\label{eq:variance_veloctiy}
        \langle v^2\rangle=\int{p(v|n_0)v^2dv}.
    \end{equation}
In our model, the theoretical value of $\langle v^2\rangle$ fits the empirical data well in Fig.~\ref{fig4}{b}. $\langle v^2\rangle$ is composed of an exponential decay and/or a power-law decay with power-law exponent -2 for different limits on $n_0$ summarized in the supplementary section~(Supplementary Section S4). We again note the predictions from the CS and KSTT models are insufficient to fully explain this observation.  In the CS model $\sigma^2(\Delta v)\propto n_0^{-2}$ as in our model however the fit is poor and in the KSTT model the conditional probability $p(v|n_0)$ is independent of $n_0$ giving an approximately constant result.

Another point of distinction for our model is the correlation between velocity and total change of order volume. In our model, the correlation. Fig.~\ref{fig4}{{c}} shows that $\langle v,\Delta n\rangle$ decreases from positive to negative for bid order and vice-versa for ask orders and both go to $0$ as $x\to\infty$ in agreement with the empirical data.  The previous works capture only the properties of order-book dynamics in certain regimes. The KSTT model assumes all of the investors are high-frequency traders, which enlarges the influence of trend-following in the region far away from price leading to a correlation that does not taper to zero far away from the trading price.  In contrast, the CS model completely ignores traders' reaction to the changing price velocity, leading to the deviation from an empirical value near the price.  Both of the previous works also neglect the influence of market activity therefore, $\langle\Delta n^2\rangle^{1/2}$ is approximately equal at different velocities while the curve our model produces agrees with the empirical data~(Fig.~\ref{fig4}{d}).

To conclude, the above simulation results and analysis indicate that our model can precisely capture and potentially explain the power law decay of the price return distribution found to be universal across a wide range of markets.  We also report the success of our model in demonstrating some of the key features of order-book dynamics as an improvement over previous work.  However one obvious limitation of our model is the lack of of temporal correlation in the traders' reactions. It is reported that the time series constructed by assigning the value $+1$ to incoming buy orders and $-1$ to incoming sell orders exhibits long memory on the Paris Bourse~\cite{bouchaud2004fluctuations}. Since the order placement $\Delta n^{OP}$ in our model follows a stable distribution which leads to the absence of long memory in the order flow, we cannot predict these results.

Finally, we point out that our model can be used for price prediction provided quality data. Current methods of predicting price, such as ARMA~\cite{box2015time} and GARCH~\cite{bollerslev1986generalized}, are based on the price series and while order book data can be over-valued in some financial analyses, our model would constitute the basis for a complementary approach to more conventional methods.  Since our model is only concerned with the mesoscopic details of order book trading, many of the complications of using order book data for price return prediction aren't an issue such as so-called iceberg orders wherein a single market maker tries to sell a large amount of a security secretly by not listing it all at once.  As long as the orders follow one-sided stable distributions and general trader reaction trends, our model is applicable.

\section*{Acknowledgements}
The authors thank Hao Zhou for helpful comments. X.S was supported by the National Natural Science Foundation of China under award numbers 61802370. X.C was supported by the National Natural Science Foundation of China under award numbers 60873245. H.S was supported by the K.C. Wong Education Foundation.
\section*{Author contributions}
C.Song conducted the project.  F.S. collected and curated datasets, and performed numeric simulation. C.S., F.S., N.A., S.H. developed the model and calculated analytical results. C.S., F.S., N.A., S.H., X.S., J.G., L.X., H.S., X.C., and N.J. analyzed data and contributed to the writing of the manuscript.
\section*{Competing Interests}
The authors declare no competing interests.

\newpage

\bibliography{bitcoinref}
\newpage
\end{document}